\documentstyle[aps,multicol,epsfig,psfig]{revtex}   

\newcommand{\beq}{\begin{equation}}  
\newcommand{\eeq}{\end{equation}}  
\newcommand{\bea}{\begin{eqnarray}}  
\newcommand{\eea}{\end{eqnarray}}  
\begin{document}     
\draft     
\title{Non-sequential triple ionization in strong fields}
\author{
Krzysztof Sacha$^{1,2}$ and Bruno Eckhardt$^1$
}     
\address{$^1$ Fachbereich Physik, Philipps Universit\"at    
	 Marburg, D-35032 Marburg, Germany}    
\address{$^2$ Instytut Fizyki im. Mariana Smoluchowskiego,
Uniwersytet Jagiello\'nski, ul. Reymonta 4,
PL-30-059 Krak\'ow, Poland}
\date{\today}     
\maketitle{ }   
  
\begin{abstract}
We consider the final stage of triple ionization of atoms in a 
strong linearly polarized laser field. 
We propose that for intensities below the saturation value for
triple ionization the process
is dominated by the simultaneous escape of three electrons
from a highly excited intermediate complex. We identify within
a classical model two pathways
to triple ionization, one with a triangular configuration of 
electrons and one with a more linear one. Both are saddles in
phase space. A stability analysis indicates that the triangular
configuration has the larger cross sections and should be 
the dominant one. Trajectory simulations 
within the dominant symmetry subspace reproduce the experimentally observed 
distribution of ion momenta parallel to the polarization axis.  
\end{abstract}     
\pacs{32.80.Fb, 32.80.Rm, 05.45.Mt}    
 
\begin{multicols}{2}  

\narrowtext

\section{Introduction}
Multiple ionization is a fundamental process in
laser-atom interactions \cite{silap1,silap2}. 
It received particular attention
when it was realized \cite{lhuiller} that the cross section was much larger
than could be expected for independent electrons model \cite{silap1,silap2}. 
One, therefore, came to distinguish two modes of multi-ionization, 
a sequential ionization processes that is
compatible with an independent electron model, and a
non-sequential one, where correlations in electron dynamics
are important. The way interactions come in is a matter of debate but the 
rescattering model \cite{Kuland,Corkum} 
has a fascinating combination
of quantum tunneling and classical dynamics of electrons
in a field and has the strongest support from experimental observations,
theoretical analyses and numerical simulations  
\cite{silap1,silap2,lhuiller,Kuland,Corkum,Kulander3,weber1,weber2,weber3,weber4,rottke,becker1,becker2,Kulander5,becker3,becker4,kopold,ullrich,lein,chen}. 
One consequence of the rescattering
model is that the final decay towards multiple ionization
is preceeded by the formation of a highly excited complex
of electrons close to the nucleus. This compound states forms during 
the collision of the rescattered electron with the core. In our analysis we
take this state as the initial condition for the final decay towards 
a multiply ionized atom.

The way two electrons escape from this compound state in non-sequential double
ionization has 
been elucidated in a recent set of experiments by Weber et al \cite{weber3}. 
They found that the electrons escape preferentially with the 
same momenta.
This observation prompted us to analyze the classical pathways
that could lead to double ionization \cite{Eckhardt,sacha}. In an extension
of Wanniers analysis \cite{Wannier,Rau,rost,ES,Rost} 
we found that for two electron
escape most classical paths lie near a subspace
of electrons moving with the same distance from the nucleus
but reflection symmetric with respect to the field axis.
Calculations of ion momenta parallel and perpendicular to
the field within this subspace \cite{Eckhardt,sacha} are in favorable agreement
with experiment. 

At these laser intensities also more than two electrons can escape 
simultaneously. In particular, triple ionization has been investigated
by Moshammer et al. \cite{rottke} and the recoil ion momentum distributions
have been measured.
It is our aim here to present the extension of the previous arguments 
\cite{Eckhardt,sacha} to the case of triple 
ionization. We identify saddles for non-sequential three electron escape 
and the relevant symmetry subspaces in which the saddles are situated.
Numerical simulations within the subspaces allow us to obtain the ion momenta 
that result from simultaneous triple ionization.
Specifically, in section II we discuss the pathways that
lead to triple ionization, identify two saddle configurations
and analyze their stability properties. In section III we then
present numerical results for the distributions of ion momenta
within these two subspaces and compare to the experimentally
observed distributions. Finally, in section IV we draw conclusions
and given an outlook to higher multiple ionization.

\section{Pathways to non-sequential triple ionization}

\begin{figure}[hbt]
\centering{\psfig{file=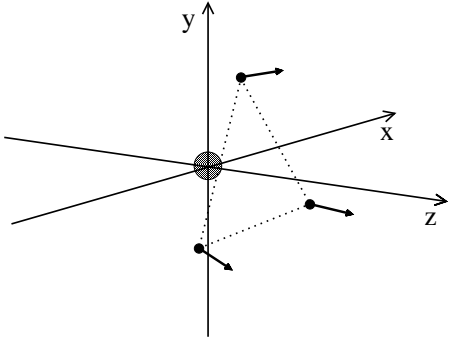,scale=0.34,angle=-90}}
\caption[]{Schematic representation of electron motion in the $C_{3v}$
symmetry subspace. The electrons are placed in a plane 
perpendicular to the field polarization axis at the vertices
of an equilateral triangle with the field axis in the center.
The circle at the origin of the coordinate system marks
the position of the nucleus.
}
\label{triangle1}
\end{figure}

As in the case of double ionization we can assume that the electron
dynamics during the final decay is fast compared to the changes
in the field \cite{Eckhardt,sacha}. Then
the straightforward extension of the double ionization analysis 
\cite{Eckhardt,sacha}
to three electron escape calls for an investigation of stationary
configurations of three electrons in an atom exposed to a strong 
static field. Keeping a high symmetry between
the electrons an obvious candidate is a configuration
where electrons are placed at the vertices of an equilateral 
triangle whose plane is perpendicular 
to the field axis (see Fig.~\ref{triangle1}). 
Deviations from this plane, in particular displacements 
in field direction, will again be amplified by electron repulsion, 
thus leading to electrons being pushed towards the nucleus and not to 
triple ionization. This configuration has $C_{3v}$ symmetry
and will be analyzed in subsection \ref{c3v}.

Further investigations show, however, that there is a second
stationary configuration. This second configuration 
has all electrons in a plane with one electron on the field axis 
and the others symmetric with respect to it. This configuration is 
of symmetry $C_{2v}$ and will be analyzed in subsection \ref{c2v}. 
The decay signatures of both configurations are different and 
the experimental observations can be used to decide between them.

In both cases we start from the Hamiltonian for three electrons
in a neutral atom in a laser field polarized along the $z$-axis.
Within dipole approximation, with infinitely heavy nucleus
and in atomic units, the Hamiltonian is
\beq
H=\frac{1}{2}({\mathbf p}_1^2+{\mathbf p}_2^2+{\mathbf p}_3^2)
+V({\mathbf r}_1,{\mathbf r}_2,{\mathbf r}_3,t),
\label{hamfull}
\eeq
with the potential energy
\bea
V&=&-\frac{3}{r_1}-\frac{3}{r_2}-\frac{3}{r_3} \nonumber\\
&&+\frac{1}{|{\mathbf r}_1-{\mathbf r}_2|}
+\frac{1}{|{\mathbf r}_2-{\mathbf r}_3|}
+\frac{1}{|{\mathbf r}_1-{\mathbf r}_3|}\nonumber\\[0.1em]
&&-(z_1+z_2+z_3)F f(t)
\label{ps}
\eea
and the time dependence of the pulse
\beq
f(t)=\sin^2(\pi t/T_d) \cos(\omega t+\phi).
\label{envel} 
\eeq
$F$, $T_d$, $\omega$ and $\phi$ stand for peak amplitude, duration, 
frequency and phase of the external field, respectively.

The projection of total angular momentum onto the polarization axis is 
conserved so that the number of degrees of freedom is reduced by one, 
leaving eight plus the time dependence from the external field. 
We work in a subspace where this component vanishes. 

Before we proceed with our analysis note a
scaling symmetry of the classical Hamiltonian (\ref{hamfull}) 
that can be used to eliminate 
one parameter. If the variables are rescaled according to
\bea
H&=&F^{1/2} H' \nonumber\\
{\bf r}&=&F^{-1/2} {\bf r}' \nonumber\\
{\bf p}&=&F^{1/4} {\bf p}' \nonumber\\
t&=&F^{-3/4} t' \nonumber\\
\omega&=&F^{3/4} \omega' 
\label{sc}
\eea
the dynamics becomes independent of the peak field amplitude, i.e. 
the system is described by the Hamiltonian (\ref{hamfull}) with $F=1$.
We will use this scaling, but drop the prime to denote the scaled
variables, for the analysis of the saddles, but we will keep the 
full field dependence for the dynamical simulations.

\subsection{$C_{3v}$ symmetry subspace}
\label{c3v}

For zero total angular momentum projection on the polarization axis,
the $C_{3v}$ symmetric configuration corresponds in cylindrical 
coordinates to electron positions given by
$z_i=Z$, $\rho_i=R$ and $\varphi_i=2\pi i/3$ with conjugate
momenta $p_{z_i}=p_Z/3$, $p_{\rho_i}=p_R/3$ and $p_{\varphi_i}=0$,
respectively. The Hamiltonian of the system in the $C_{3v}$
subspace then reads
\beq
H(p_R,p_Z,R,Z,t)=
\frac{p_R^2+p_Z^2}{6}+V(R,Z,t),
\label{h}
\eeq
with potential energy
\beq
V=-\frac{9}{\sqrt{R^2+Z^2}}+\frac{3}{2R\sin (\pi/3)}
-3Zf(t)
\label{p}
\eeq
The three terms in Eq.~(\ref{p}) are the cumulative interaction of 
three electrons with the triply-charged nucleus, 
the repulsion energy between electrons at distances $2R\sin(\pi/3)$ 
and the interaction with the external field, respectively. 
The variables are scaled
according to (\ref{sc}) so that the field amplitude $F$ is absent.

\begin{figure}[hbt]
\centering{\psfig{file=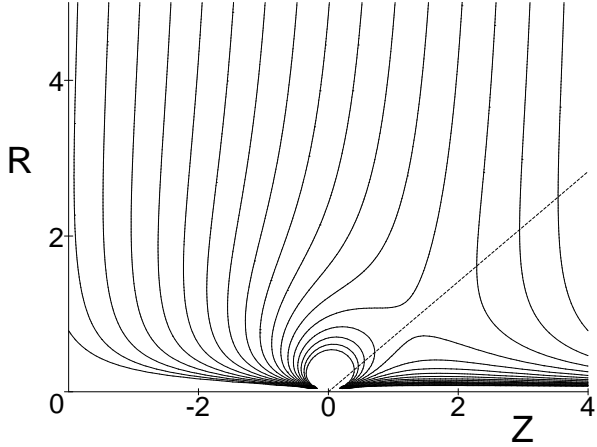,scale=0.34, clip=true,angle=-90}}
\caption[]{
Equipotential curves of the potential energy (\protect{\ref{p}})
for three electrons 
The potential is plotted for a field strength $F f(t)=1$. The
saddles move along the dashed lines when the electric field points in the
positive $z$ direction and along the one obtained by reflections on $z=0$
during the other half of the field cycle. 
}\label{one2ab}
\end{figure}

Electron motion close to the nucleus is much faster than the change of 
the phase of the laser field applied in the experiments
\cite{weber1,weber2,weber3,weber4,rottke,Eckhardt,sacha}. Therefore,  
we can use an adiabatic approximation and 
gain insight into the qualitative features of the ionization process 
by analyzing
the potential (\ref{p}) for fixed external field. Note, however, 
that we use the full time dependence for the determination
of the final ion momenta below.
The potential (\ref{p}) for a given time and thus a fixed value of
$f(t)$ is shown in Fig.~\ref{one2ab}. The saddle is located along
the lines $Z_s=r_s\cos\theta_s$ and 
$R_s=r_s\sin\theta_s$ with $\theta_s=\theta$ 
or $\theta_s=\pi-\theta$ where
\beq
\theta=\arctan\frac{1}{\sqrt{2}}\approx 35^\circ
\label{th}
\eeq
and
\beq
r_s^2={\sqrt{6}}/{|f(t)|} \,.
\label{rs}
\eeq
The energy of the saddle is 
\beq
V_s=-6^{3/4}2\sqrt{|f(t)|}.
\label{se}
\eeq
During a field cycle the saddle moves in from infinity along the line
$\theta_s=\theta$, back out to infinity and then in and out again along
the line $\theta_s=\pi-\theta$. 

For the dynamics in full phase space the stability properties of 
the configuration are important. Since the electrons move fast compared
to the frequency of the field we can for this analysis employ
an adiabatic approximation and work with a constant field. Then we can
scale out the field amplitude completely and put $f(t)=1$.
The triangular electron configuration then becomes stationary and
the dynamics near this saddle is governed by the eigenvalues of 
the linearization. Within the $C_{3v}$ symmetry subspace 
the triangular electron configuration has one unstable direction
corresponding to a crossing of the saddle. 
In these scaled units the Lyapunov exponent for motion along
this reaction coordinate \cite{Wigner,Pollak} is $\lambda_{r}=1.1054$.
The other eigenmode of the saddle in the subspace is stable.

Harmonic approximation in the full eight-dimensional configuration 
space results in three additional pairs of degenerate eigenvalues. 
Of these pairs two are stable and one is unstable. The unstable eigenspace
points towards asymmetric configurations, 
has a Lyapunov exponent $\nu_a=1.4496$ and is spanned by the vectors
\bea
\rho_1=R_s+w_a,\ & z_1=Z_s+2.0642w_a, \cr
\rho_2=R_s+w_a,\ & z_2=Z_s+2.0642w_a, \cr
\rho_3=R_s-2.0000w_a,\ & z_3=Z_s-4.1284w_a,
\label{dic1}
\eea
and
\bea
\rho_1=R_s-u_a,\ & z_1=Z_s-2.0642u_a, \cr
\rho_2=R_s+u_a,\ & z_2=Z_s+2.0642u_a, \cr
\rho_3=R_s,\ & z_3=Z_s.
\label{dic2}
\eea
The first mode describes a motion where 
for positive $w_a$ two electrons move away from
and one towards the nucleus or vice versa for 
negative $w_a$. The other mode leaves one electron on the saddle
and brings one closer and one further out.

One may compare the period of the laser field applied in the experiments 
\cite{rottke} with  the
time scales for crossing of the saddle in the $C_{3v}$ subspace and 
for departure from the subspace. For peak field intensity 
$1.5\cdot 10^{15}$~W/cm$^2$ and 
for wavelength of 795~nm, the field period 
(in the scaled variables (\ref{sc}))
is $2\pi/\omega=33.6$ while $1/\lambda_r=0.9$ and $1/\nu_a=0.69$.
This indicates that crossing the saddle and departure from the
non-sequential triple ionization manifold take place rather quickly, 
justifying the adiabatic analysis a posteriori.

\subsection{$C_{2v}$ subspace}
\label{c2v}

\begin{figure}[hbt]
\centering{\psfig{file=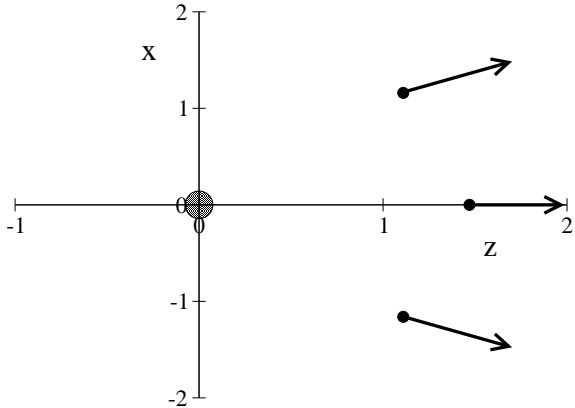,scale=0.34, clip=true,angle=-90}}
\caption[]{
Saddle in the $C_{2v}$ symmetry subspace. The three electrons are placed 
at the stationary point of the potential ({\protect \ref{hamc2v}}) 
(for $Ff(t)=1$). The arrows indicate
the unstable mode of the saddle corresponding to the simultaneous escape 
of all electrons. The circle at the origin of the coordinate
system marks the position of the nucleus.
}\label{sadd4}
\end{figure}

In addition to the triangular configuration there is one
with all electrons in a plane. For fixed external 
field the potential (\ref{ps}) possesses a stationary point
with one electron on the polarization axis and the other two 
placed on either side symmetrically with respect to the field axis
(Fig.~\ref{sadd4}). Without loss of generality we may assume that the
electrons are confined to the $x$-$z$-plane and then the $C_{2v}$
subspace is spanned by 
\bea
y_1&=&y_2=y_3=0\cr
p_{y_1}&=&p_{y_2}=p_{y_3}= 0\,,\cr
x_1&=&0\,,\quad p_{x_1}=0\,,\cr
x_2&=&-x_3=x\,,\quad p_{x_2}=-p_{x_3}=p_{x}/2\,, \cr
z_2&=&z_3=z\,,\quad  p_{z_2}=p_{z_3}=p_z/2\,.
\eea
The phase space for motion in this subspace is six-dimensional, with
variables $(x,z,z_1, p_x, p_z, p_{z1})$. The Hamiltonian becomes
\bea
H&=&\frac{p_x^2+p_z^2}{4}+\frac{p_{z_1}^2}{2}-\frac{6}{\sqrt{x^2+z^2}}
-\frac{3}{z_1} \cr
&& +\frac{1}{2|x|}+\frac{2}{\sqrt{x^2+(z-z_1)^2}} \nonumber\\[0.2em]
&& -(2z+z_1)f(t).
\label{hamc2v}
\eea
In contrast to the $C_{3v}$ subspace the saddle cannot be given
analytically and has to be found numerically. For 
$f(t)=1$, the saddle is located at $|x_s|=1.1607$, $z_s=1.1143$ 
and $z_{1,s}=1.4665$ and has the potential energy
$V_s=-7.3902$. The configuration is elongated along a 
line perpendicular to the field axis, with axial position of the 
central electron further out than that of the outer two. 
Nevertheless, the distance of the central electron to the nucleus 
is smaller than that of the outer electrons. It thus experiences a 
stronger attractive force and the balance of force requires 
that the other two electrons contribute a component pointing 
away from the nucleus, i.e. that $|z_s|<|z_{1,s}|$, as indeed observed.

Harmonic approximation in the $C_{2v}$ subspace around the saddle
reveals one stable and two unstable modes. One of the unstable directions, 
with Lyapunov exponent $\lambda_{2r}=1.0980$, corresponds to simultaneous 
escape of three electrons: in the eigenspace
\bea
z_1&=&z_{1,s}+w_{2r} \cr
x&=&x_s+0.6183w_{2r} \cr
z&=&z_s+1.1417w_{2r},
\eea
all three components increase for positive $w_{2r}$. This eigenspace
thus corresponds to the reaction coordinate in the triangular configuration.
The other eigenmode 
\bea
z_1&=&z_{1,s}+u_{2a} \cr
x&=&x_s-0.2313u_{2a} \cr
z&=&z_s-0.3127u_{2a},
\eea
with Lyapunov exponent $\nu_{2a}=1.7937$ is related to double or single 
ionization, according to the sign of $u_{2a}$.

Harmonic approximation in the full space gives in addition to the
modes within the subspace three 
stable modes, two unstable ones reflecting divergence from the $C_{2v}$ 
subspace and one neutral mode connected with an overall rotation around the
field axis. The neutral mode is connected with conservation
of angular momentum and was eliminated in the $C_{3v}$ case by
transformation to polar coordinates; it shows up in the $C_{2v}$ analysis
where cartesian coordinates are more convenient since
one electron would come to lie on the singularity of the polar
coordinate system.
The first unstable mode with Lyapunov exponent $\nu_{2b}=0.9024$ 
corresponds to deviations along the $y$-axis,
\bea
y_1&=&u_{2b} \cr
y_2&=&-0.5752u_{2b} \cr
y_3&=&-0.5752u_{2b} \,.
\eea
The other unstable mode, with the Lyapunov exponent $\nu_{2c}=1.3712$,
is related to symmetry breaking within the $x$-$z$-plane,
\bea
\matrix{
x_1=0.0746u_{2c} & \quad z_1=z_{1,s} \cr
x_2=x_s+0.6598u_{2c} & \quad z_2=z_s+u_{2c} \cr
x_3=-x_s+0.6598u_{2c}& \quad z_3=z_s-u_{2c}\,.
}
\eea
The Lyapunov exponents are of similar magnitude as in the 
$C_{3v}$ case and the 
adiabatic approximation remains justified.

\subsection{Comparison between the subspaces}
Analysis of the stationary points in phase space shows that 
there are two pathways to non-sequential triple ionization.
The triangular configuration has a potential barrier 
(for scaled field $Ff(t)=1$) of $-7.6673$, slightly lower
than the one $-7.3902$ for the saddle in the $C_{2v}$ subspace.
Therefore, for increasing energy in the compound complex, 
triple ionization via the triangular path becomes possible first.
The triangular state is also less unstable than the 
$C_{2v}$ configuration: it has a largest Lyapunov exponent for 
symmetry breaking perturbations of $\nu_a=1.4496$, 
as compared to $\nu_{2a}=1.7937$. For motion along
the reaction coordinates the Lyapunov exponent of the 
triangular configuration is $\lambda_r=1.1054$ and thus slightly larger
than $\lambda_{2r}=1.0980$ for the $C_{2v}$ configuration.
The relative phase space weight of trajectories near the 
symmetry subspaces is determined by a competition between
motion along the reaction coordinate and amplification of
deviations from symmetry 
\cite{ES,Rost}. Generally, the faster the motion
along the reaction coordinate and the slower the symmetry
breaking the larger the phase space that is dominated
by the saddle.

There are no quantitative estimates for the phase space
region influenced by the saddle, except close to a threshold,
where a suitable extension of Wanniers arguments can be
employed \cite{Wannier,Rau,rost}. 
For triple ionization
and in the presence of several symmetry breaking modes, the 
generalization \cite{ES,Rost} gives a threshold behavior
$\sigma(\Delta E) \propto (\Delta E)^\alpha$ with exponent
\beq 
\alpha = \left.\left(\sum \nu_i\right)\right/\lambda_j
\eeq
where $\lambda_j$ is the Lyapunov exponent of the reaction
coordinate and the $\nu_i$ are the ones of the symmetry breaking modes.
Specifically, for our example here we find for the triangular configuration
\beq
\alpha_3=2.6228
\eeq
and for the $C_{2v}$ configuration
\beq
\alpha_2=3.7043\,.
\eeq
This already suggests that the triangular configuration will be
the more dominant configuration. Besides the threshold behavior
the two saddles also differ considerably in the final momentum
distributions of ionizing electrons and this will be analyzed in the next section.

\section{Ion momenta distributions}

In the present section we show numerical simulations of the 
non-sequential ionization. We restrict
our calculations to the symmetry subspaces and compare the
resulting final ion momentum distributions. Even though the symmetry
subspaces are of zero phase space weight and are furthermore unstable
non-sequential triple ionization trajectories in full phase space have 
to pass sufficiently close to the saddles and 
these subspaces so that they will show distributions
very close to the ones obtained in the subspaces \cite{Eckhardt,sacha}. 
We can,
therefore, also compare our data with experimental distributions.
In order to allow comparison with
experiments we now omit the elimination of the maximal
field amplitude $F$ by the rescaling (\ref{sc}).

In the experiments of Moshammer et al. \cite{rottke} on triple ionization 
of Ne ultrashort (30~fs) laser pulses at 795~nm wavelength 
and with peak intensities of 1.5~PW/cm$^2$ were used. This corresponds
to a frequency of $\omega=0.057$~a.u. (atomic units), a 
pulse duration, measured as full width at half maximum, 
of 11 periods $2\pi/\omega$ and a maximal field strength
of $F=0.207$~a.u.. Among the measured data we focus on
the distributions of ion momenta parallel 
to the polarization axis. In the limit of negligible small 
momentum transfer by the absorbed photons, the ion momentum $\vec p_{ion}$ 
reflects the sum of the momenta of the emitted electrons, 
$\vec p_{ion}=-\sum_i \vec p_i$ \cite{weber1,rottke}. 

The three-electron Hamiltonian (\ref{hamfull}) corresponds to 
lithium. In order to relate the calculations to other atoms with
more electrons interactions with core electrons have to be
neglected 
and the point of reference in energy has to be
shifted to the threshold for triple ionization.
Specifically, for the modeling of the experiments 
\cite{rottke} on triple ionization in Ne we thus assume that in a rescattering 
process the energy transfer is less than the threshold for triple ionization 
(about $4.6$~a.u.).
The precise value of the energy transfered in the rescattering event
and thus of the energy $E$ of the initial compound state cannot be 
determined within our model and constitutes a free parameter.
Similarly, the time $t_0$ during the pulse when the 
rescattering complex forms is a second free parameter. 
But both parameters can be determined rather reliably by
comparison with experiments \cite{sacha}. We now turn to the simulations
within the two symmetry subspaces.

\begin{figure}[hbt]
\centering{\psfig{file=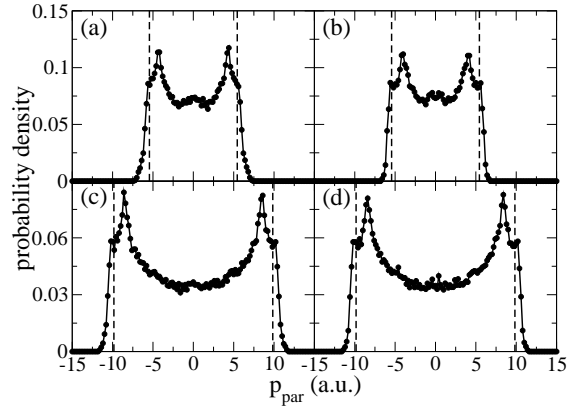,scale=0.3, clip=true,angle=-90}}
\caption[]{
Final distributions of parallel ion momenta calculated in the $C_{3v}$
symmetry subspace for the initial energy $E=-0.5$~a.u., peak field amplitude
$F=0.207$~a.u., and pulse duration $T_d=20\times 2\pi/\omega$, where
$\omega=0.057$~a.u.. The initial times under the pulse envelope 
are $t_0=0.25T_d$ (a), $t_0=0.75T_d$ (b), $t_0=0.4T_d$
(c), and $t_0=0.6T_d$ (d). Dashed lines indicate the estimates
$\pm3Ff_p(t_0)/\omega$.
Note that the distributions are essentially the same independently if one 
chooses $t_0$ before or after the peak field value provided $f_p(t_0)$ is the 
same. The distributions are based on about $8\cdot10^4$ trajectories. 
}\label{time5}
\end{figure} 

\subsection{Simulations in the C$_{3v}$ subspace}

In a first series of simulations we focus on the triangular
configuration. We fix the initial energy $E$ 
and analyze the dependence of the results on the initial time $t_0$. 
To this end we chose a microcanonical distribution of initial 
conditions in the $C_{3v}$ symmetry subspace for energy $E=-0.5$~a.u. 
and random phases $\phi$ for different initial time $t_0$. 
The final distributions of ion
momenta parallel to the polarization axis are shown in Fig.~\ref{time5}. 
As in our previous analysis of the non-sequential double ionization 
process \cite{Eckhardt,sacha} 
the distributions has a double hump structure and the width of the 
distributions increases when the initial time $t_0$ approaches
the peak amplitude of the field. The
width of the distributions can be
estimated from the maximal energy a free electron can acquire in
the field, i.e. twice the ponderomotive energy. For three 
electrons ejected in the same direction, 
the corresponding maximal ion momentum depends on the amplitude of 
the field at the point in time when they are ionized, i.e.
\beq
p_{max}=\frac{3Ff_p(t)}{\omega}
\label{pmax}
\eeq 
where $f_p(t)=\sin^2(\pi t/T_d)$ is the pulse envelop. 
If this time is taken to equal the starting time $t_0$ of the simulations
we find values of $p_{max}$ which 
correspond very well to the widths of the distributions in Fig.~\ref{time5}.
Fig.~\ref{time5} indicates also that the distributions are basically the same
independent of whether $t_0$ is chosen before or after the maximum of the 
pulse, provided $f_p(t_0)$ is the same. This implies that the dominant 
ionization takes place 
during the first field cycle after formation of the complex. 
Fitting the width of 
the calculated distribution to the experimental results allows one
to estimate the moment in the pulse when the majority of the triply 
ionized ions are created.

\begin{figure}[hbt]
\centering{\psfig{file=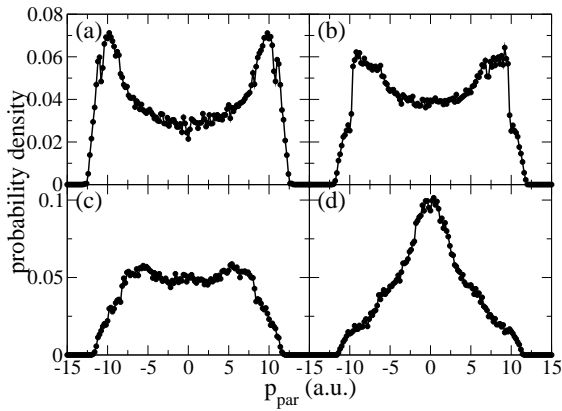,scale=0.3, clip=true,angle=-90}}
\caption[]{
Ion momentum distributions in the $C_{3v}$ subspace for fixed
initial time $t_0=0.5T_d$ and different initial energies
$E=-0.05$~a.u. (a), $E=-1.5$~a.u. (b), $E=-2$~a.u. (c), and
$E=-3$~a.u. (d). The parameters are the same as in 
Fig.~{\protect \ref{time5}}. The minimal energy of the saddle, 
obtained for the maximal field amplitude, is $-3.49$~a.u..
}\label{en6}
\end{figure}
The dependence of the ion momentum distributions on the initial 
energy $E$ for fixed initial time $t_0$ is presented in Fig.~\ref{en6}. 
The widths of the distributions does not change significantly with 
$E$ but their shape does. For initial energy close to the minimal 
saddle energy, which for $F=0.207$~a.u. is $-3.49$~a.u., 
the distributions show a single maximum only while for higher energies
two humps form. The electrons that cross the saddle when 
the energy is near the saddle energy are slow and the combined interactions 
of the external and Coulomb fields shapes the distribution \cite{sacha}. 
The situation is different for the high energy case: then shortly 
after the electrons cross the saddle the interaction with the 
electric field is stronger than the attraction to the nucleus 
and the distribution is mostly shaped by the laser 
field \cite{sacha}. As the initial energy of the complex is determined in 
a rescattering event and is higher for stronger external fields, 
the distribution of ion momenta should show a transition from
a distribution with a single maximum near the threshold for  
triple ionization to one with a double maximum higher up. For
even higher fields the constraints from the triangular configuration
are relaxed and less symmetric modes of triple ionization become possible.

\subsection{Numerical simulation in the $C_{2v}$ subspace}

We now turn to the ion momentum distributions in the 
$C_{2v}$ subspace. Much of the analysis proceeds as in 
the previous subsection, except for the difference
in Hamiltonians (here it is (\ref{hamc2v})) and the
choice of initial conditions. The microcanonical ensemble 
of the initial conditions is not straightforward to realize because 
(even for fixed time) 
for nonzero external field the system is open. In the
previous section, the difficulties could be overcome by
constraining initial conditions to lie on the energy shell 
and in the hypersurface $Z=0$ (for details see \cite{sacha}). 
Here such a restriction is not sufficient, phase space remains open.
The troublesome configurations are those where electron distances
to the nucleus are strongly asymmetric, one being close and two far
away or vice versa. Since we assume that the initial state is 
formed during a rescattering and should be confined to region close
to the core, such configurations cannot be formed. We,
therefore, choose initial
conditions microcanonically and require in addition that distances of the
electrons to the nucleus are not larger than the minimal distance of the 
saddle at maximal field to the nucleus. 

\begin{figure}[hbt]
\centering{\psfig{file=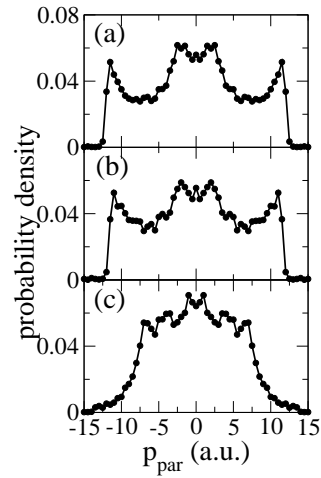,scale=0.35, clip=true,angle=-90}}
\caption[]{
Final distributions of parallel ion momenta calculated for non-sequential triple
ionization in the $C_{2v}$ symmetry subspace for the peak field amplitude
$F=0.207$~a.u., pulse duration $T_d=20\times 2\pi/\omega$ (where
$\omega=0.057$~a.u.) and initial time in the pulse duration $t_0=0.5T_d$. 
The initial energy is $E=-0.05$~a.u. (a), $E=-0.5$~a.u. (b), $E=-2$~a.u. (c)
--- the minimal energy of the saddle is $-3.36$~a.u..
The distributions are obtained from a few thousands trajectories. 
}\label{c2v8}
\end{figure}
As in the previous section, the width of the momentum distributions 
determines the initial time $t_0$ when triple ionization takes place.
The dependence on initial energy $E$ is also similar, with a 
collapse of all structures into a single peak near zero momentum
as the initial energy decreases towards the energy of the saddle
(Fig.~\ref{c2v8}). 

\begin{figure}[hbt]
\centering{\epsfig{file=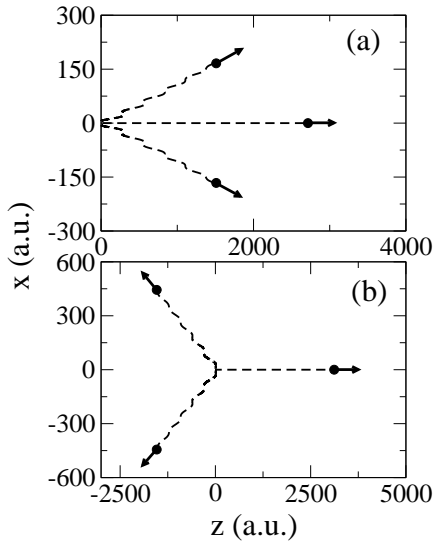,scale=0.38, clip=true,angle=-90}}
\caption[]{
Examples of trajectories corresponding to simultaneous three electron 
escape by crossing the saddle in the $C_{2v}$ symmetry subspace. 
The electrons, when crossing the saddle, move initially 
towards the positive $z$ direction (the saddle in both cases is
located at $z>0$). However, depending on the phase of 
the field, their final momenta can point in either the same (a) or 
opposite (b) direction.
}\label{tr9}
\end{figure}
In Fig.~\ref{c2v8} we show final parallel ion momentum distributions 
corresponding to
non-sequential triple ionization in the $C_{2v}$ subspace for three different
initial energy $E$. The structure of the distributions is 
significantly different from that in the previous section.
In Fig.~\ref{c2v8}(a) and \ref{c2v8}(b), besides two local maxima on 
the edges of
distributions, a strong contribution zero momentum value that is split 
into two maxima appears. The origin of these structures can be 
related to qualitatively different electron paths. Initially, after
crossing the saddle all three electrons escape towards
the same side of the nucleus. However, the final momenta 
along the polarization axis may point to either the same or 
opposite directions, depending on 
the initial conditions and the phase of the field, 
as shown in Fig.~\ref{tr9}. 
If the final momenta point in the same direction, they can be 
much larger than if they point in opposite directions.
Therefore, the edge maxima are the effect of the electrons escape towards 
the same direction while the central structure 
corresponds to an escape in opposite directions. 

\subsection{Comparison with experiment}

\begin{figure}[hbt]
\centering{\psfig{file=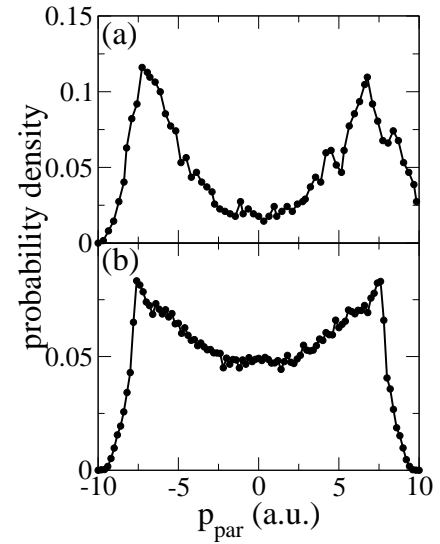,scale=0.4,angle=-90}}
\caption[]{
Comparison between experimental (a) and numerical (b) ion momenta.
Ion momentum distributions are from the experiment of \cite{rottke}
on triple ionization of Ne atoms in the focus of 795~nm, 30~fs (i.e. about
$11\times 2\pi/\omega$) laser pulses at peak intensity of 
$1.5$~PW/cm$^2$ (i.e. peak field strength $F=0.207$~a.u.).
The numerical distributions are for the $C_{3v}$
symmetry subspace with initial energy $E=-1$~a.u. and $t_0=0.33T_d$ where
$T_d/2=11\times 2\pi/\omega$, see Eq.~{\protect (\ref{envel})}.
}\label{exp7}
\end{figure}

Given the qualitatively very different distributions of final ion momenta
in the two subspaces we can use a comparison with experimental data 
\cite{rottke} to draw conclusions about the dominant ionization mode. 
In Fig.~\ref{exp7} we show the
experimental parallel ion momentum distribution together with the one
calculated from the $C_{3v}$ symmetry subspace. 
In the calculation we have chosen the initial time $t_0=0.33T_d$ that gives
the width of the resulting distribution in agreement with the experiment. The
initial energy has been taken as $E=-1$~a.u. but it can be taken in 
a wide range of values above $-1.5$~a.u. without significant changes
in momentum distributions (Fig.~\ref{en6}).
The main features of the experimental distribution
are well reproduced in our calculations. However, the minimum in the 
distribution is less pronounced than that observed in the experiment.
Overall, we take this good agreement as strong indication that triple 
ionization occurs in the neighborhood of the symmetric process discussed here. 

The distribution for perpendicular ion momenta, which have been measured in the
experiment \cite{rottke}, can not be calculated within the symmetry subspace
since this component vanishes exactly by symmetry. In full phase space
ionizing electrons pass close to the saddle but not exactly symmetrically and
consequently the transverse ion momenta do not vanish.

\section{Conclusion}

We have considered the pathways along which a highly excited three
electron complex can decay towards triple ionization. In non-sequential
multiple ionization of atoms in strong laser fields such a complex
forms when the rescattered electron collides with the core. The
subsequent non-sequential decay then has to follow the pathways 
discussed here. We found two possible pathways, one proceeding
via a symmetric triangular configuration, and one via a 
planar elongated configuration. A stability analysis shows that 
the ionization close to the $C_{3v}$ symmetry subspace occurs
at lower field intensities and has a larger cross section
(close to threshold). 
Trajectory simulations within this subspace give the final ion momentum 
distributions in very good agreement with the recent experimental data
\cite{rottke}.
The $C_{2v}$ configuration has a higher threshold and 
smaller cross section
and gives a momentum distribution different from the experimentally
observed one, so that we conclude that the triangular configuration
is indeed the dominant one.

We have not discussed pathways to sequential triple ionization, and 
one can expect that there are several, including sequential ionization of
one electron after the other, a double ionization followed by a 
single ionization or the other way around, or various
combinations of partial ionizations and rescatterings. 
As in the case of double ionization \cite{Eckhardt,sacha}
one can also have sequential ionization from failed attempts to 
triple ionize along one of the symmetric saddles.
The experimental results of
\cite{rottke} suggest that, at least in the intensity range considered here, 
correlated non-sequential electron escape dominates.
This is further supported by our previous analysis of 
double ionization \cite{sacha}, where we found that the 
double hump structure observed in the experimental ion momentum 
distributions \cite{weber1,weber2,weber3,weber4,rottke} 
does not appear in the sequential process. That is,
the electron momenta are not correlated and the resulting 
distributions show a strong single maximum instead of the 
double hump structure.

Finally, let us mention that part of our analysis can be easily extended to
non-sequential multiple ionization where the number of escaping electrons, 
$N$, is greater than three \cite{Eckhardt}. 
The triangular configuration extends
to a placement of electrons on a ring with $C_{Nv}$ in plane 
perpendicular to the field axis. Calculation of the positions and the 
energy of the saddle shows that the energy is non-monotonic and begins 
to increase with $N$ for more than 9 electrons. For more than 13
electrons the saddle does not exist any more. Therefore, for many
electrons the configurations with highest symmetry will disappear
and others will dominate the ionization process. For triple
ionization we have identified one, the planar configuration, and
for more than three electrons we expect a rapidly increasing number
of stationary configurations.
The identification of a dominant pathway will then require
an appropriate stability analysis and an application
of the generalized Wannier threshold law \cite{ES,Rost}.

\section{Acknowledgment}

We would like to thank H. Giessen, W. Becker, R. Kopold and H. Rottke 
for discussions. Financial support by the Alexander von Humboldt
Foundation and KBN under project 5~P03B~088~21 are gratefully acknowledged. 


 
\end{multicols}
\end{document}